\documentstyle[prl,aps,epsfig,amscd]{revtex}

\twocolumn

\begin{document}
\draft

\def\up{|\uparrow\,\rangle}
\def\dn{|\downarrow\,\rangle}
\def\upd{\langle\, \uparrow |}
\def\dnd{\langle\, \downarrow |}
\def\upup{|\uparrow\,\uparrow \rangle}
\def\dndn{|\downarrow\,\downarrow \rangle}
\def\updn{|\uparrow\,\downarrow \rangle}
\def\dnup{|\downarrow\,\uparrow \rangle}
\def\upupd{\langle \uparrow\,\uparrow |}
\def\dndnd{\langle \downarrow\,\downarrow |}
\def\updnd{\langle \uparrow\,\downarrow |}
\def\dnupd{\langle \downarrow\,\uparrow |}

\def\beqn{\begin{equation}}
\def\eeqn{\end{equation}}
\def\beqnar{\begin{eqnarray}}
\def\eeqnar{\end{eqnarray}}
\newcommand{\tfrac}[2]{{\textstyle\frac{#1}{#2}}}
\newcommand{\mb}[1]{\mbox{\boldmath{$#1$}}}
\def\ba{\begin{array}}
\def\ea{\end{array}}
\newcommand{\etal}{{\em et al. }}
\newcommand{\eqn}[1]{(\ref{#1})}
\newcommand{\arowspace}[1]{\renewcommand{\arraystretch}{#1}}
\newcommand{\acolspace}[1]{\renewcommand{\arraycolsep}{#1}}

\title{Quantum Simulations on a Quantum Computer}

\author{
S. Somaroo{$^1$}, 
C. H. Tseng{$^{1,2}$}, 
T. F. Havel{$^1$}, R. Laflamme{$^3$}, 
and D. G. Cory{$^4$}\footnote{
Correspondence and requests for materials should 
be addressed to D. G. C. (email:{\it dcory@mit.edu}).}
}

\address{
{$^1$}BCMP Harvard Medical School, 240 Longwood Avenue, Boston MA 02115\\
{$^2$}Harvard-Smithsonian Center for Astrophysics, 60 Garden Street, Cambridge, MA 02138\\
{$^3$}Los Alamos National Laboratory, Los Alamos, NM 87455\\
{$^4$}Department of Nuclear Engineering, Massachusetts Institute of 
Technology, Cambridge, MA 02139
}

\date{\today}

\maketitle

\begin{abstract}
We present a general scheme for performing a simulation of the dynamics of one
 quantum system using another. This scheme is used to experimentally simulate 
the dynamics of truncated quantum harmonic and anharmonic oscillators using 
nuclear magnetic resonance. We believe this to be the first explicit physical 
realization of such a simulation.
\end{abstract}

\narrowtext

In 1982, Richard Feynman proposed that a quantum system would be more 
efficiently simulated by a computer based on the principles of quantum 
mechanics rather than by one based on classical mechanics \cite{Feynman}. 
Recently, it has been pointed out that it should be possible to efficiently 
approximate any desired Hamiltonian within the standard model of a quantum 
computer by a sparsely coupled array of two-state systems 
\cite{Lloyd1,Lloyd2,Zalka}. Many of the concepts of quantum simulation are 
implicit in the average Hamiltonian theory developed by Waugh and colleagues 
to design NMR pulse sequences which implement a specific desired effective NMR
 Hamiltonian \cite{Waugh}.   Here we show the first explicit simulation of one
 quantum system by another; namely the simulation of the kinematics and 
dynamics of a truncated quantum oscillator by an NMR quantum information 
processor \cite{Cory,Gershenfeld}.  Quantum simulations are shown for both an 
undriven harmonic oscillator and a driven anharmonic oscillator.

A general scheme for quantum simulation is summarized by the following diagram:
\[
\begin{CD}
{\rm \bf Simulated (S)} & & {\rm \bf Physical (P)}\\
|s\rangle  @>\phi>>  |p\rangle \\
@V{U=e^{-i{\cal H}_sT/\hbar}}VV 		@VV{V_T}V \\
|s(T)\rangle @<\phi^{-1}<<  |{p_{\tiny T}}\rangle.
\end{CD} 
\]
The object is to simulate the effect of the evolution 
$|s\rangle \stackrel{U}{\longrightarrow} |s(T)\rangle$ using the physical 
system $P$. To do this, $S$ is related to $P$ by an invertible map $\phi$ 
which determines a correspondence between all the operators and states of $S$ 
and of $P$. In particular, the propagator $U$ maps to $V_T = \phi^{-1} U \phi$.
The challenge is to implement $V_T$ using propagators $V_i$ arising from the available external interactions with intervening periods of natural evolution 
$e^{-i {\cal H}_p^{0} t/\hbar}$ in $P$ so that 
$V_T = {\Pi}_{i} e^{-i {\cal H}_p^{0} t_i(T)}V_i$. 
If a sufficient class of simple operations (logic gates) are implementable in 
the physical system, the Universal Computation Theorem 
\cite{Shoretal,Lloyd3,Divincenzo} guarantees that any operator (in particular 
$V_T$) can be composed of natural evolutions in $P$ and external interactions.
For unitary maps $\phi$, we may write $V_T = e^{-i\overline{\cal H}_p T/\hbar}$
where $\overline{\cal H}_p \equiv \phi^{-1}{\cal H}_s\phi$ can be identified 
with the average Hamiltonian of Waugh. After 
$|p\rangle \stackrel{V_T}{\longrightarrow} |p_{T}\rangle $, the final map 
$\phi^{-1}$ takes $|p_{T}\rangle \rightarrow |s(T)\rangle$ thereby effecting 
the simulation $|s\rangle \rightarrow |s(T)\rangle$.
Note that ${\cal H}_s(T)$ can be a time dependent Hamiltonian and that $T$ is 
viewed as a parameter when mapped to $P$. This implies that the physical times
$t_i(T)$ are parameterized by the simulated time $T$.

Liquid state NMR quantum computers are well suited for quantum simulations 
because they have long spin relaxation times ($T_1$ and $T_2$) as well as the 
flexibility of using a variety of molecular samples. In particular, the 
coupling between the nuclear spins, usually dominated by the `scalar' coupling
($J$), may be reduced at will by means of radiofrequency pulses. Typically spin 1/2 nuclei are used.  Thus, the kinematics of any $2^N$ level quantum system could be simulated using a given 
$N$-spin molecule. 

We chose to simulate a quantum harmonic oscillator (QHO) with a 4 level, 
2-spin system P being the two proton nuclear spins in 2,3- dibromothiophene. 
The Hamiltonian of a QHO is
$
{\cal H}_{\rm QHO} \;\equiv\; \hbar\Omega(\hat{N} + \frac{1}{2})
 =  {\sum}_{n} \hbar\Omega(n + \frac{1}{2})|n\rangle\langle n|,
$
where $\Omega$ is the oscillator frequency, and $|n\rangle$ are the orthonormal
 eigenstates of the number operator $\hat{N}$. Since the nuclear spin 
eigenstate space is finite dimensional (4 levels, in this case), only a 
truncated version of the infinite dimensional oscillator was simulated.  
However, as noted above, for $N$ spins, there are $2^N$ levels.  A convenient 
unitary mapping, $\phi$, between the energy eigenstates of the QHO and a 
2-spin system is:
\beqn
\ba{rcccl}
|n=0\rangle & \longleftrightarrow & \up\up & \equiv & \upup \\
|n=1\rangle & \longleftrightarrow & \up\dn & \equiv & \updn \\
|n=2\rangle & \longleftrightarrow & \dn\dn & \equiv & \dndn \\
|n=3\rangle & \longleftrightarrow & \dn\up & \equiv & \dnup .
\ea
\label{map1}
\eeqn
While any of a number of mappings would suffice, this mapping is convenient 
since $\Delta n = \pm 1$ correspond to allowed transitions in $P$. This mapping generalizes to a Gray code. Also note 
that the spin basis, permuted under $\phi$, is now not in order of increasing 
energy in $P$.

When truncated, $e^{-i{\cal H}_{QHO}T/\hbar}$  is mapped onto the 2-spin 
system as follows:


\beqnar
U \;=\; e^{-i{\cal H}_sT/\hbar}\nonumber\\  
\equiv  \exp [-i (\tfrac{1}{2}|0\rangle\langle 0 | +  \tfrac{3}{2}|1\rangle\langle 1|)\nonumber\\ +\tfrac{5}{2}|2\rangle\langle 2| +\tfrac{7}{2}|3\rangle\langle 3| )\Omega T ] \nonumber\\
 \;\stackrel{\phi}{\longrightarrow}\; V_{T}\;=\;e^{-i\overline{\cal H}_p T/\hbar}\nonumber\\ 
\equiv  \exp [-i (\tfrac{1}{2}\upup\upupd +  \tfrac{3}{2}\updn\updnd\nonumber\\
+\tfrac{5}{2} \dndn\dndnd +\tfrac{7}{2}\dnup\dnupd )\Omega T ].
\nonumber
\eeqnar
\narrowtext

Using the Pauli matrices $\{\sigma_x,
\sigma_y, \sigma_z\}$ as a basis for the 2-spin density matrices 
\cite{GApaper}, we may write
\beqn
\ba{rcccl}
V_T \;=\; e^{-i\overline{\cal H}_p T/\hbar}  & = &  \exp\left[i(\sigma_z^2(1+
\tfrac{1}{2}\sigma_z^1) -2)\Omega T\right].
\ea
\label{ham1}
\eeqn
Implementing the operator \eqn{ham1} on the 2-spin system thus constitutes a 
simulation of the truncated QHO.  This is easily done by making various 
refocussing adjustments to the physical 2-spin
propagator $e^{-i{\cal H}_p^{0}t_i/\hbar}$, obtained from the natural 
Hamiltonian
$
{\cal H}_p^{0} \equiv \frac{\hbar}{2}((\omega_1-\omega_0) \sigma_z^1 + 
(\omega_2-\omega_0) \sigma_z^2 + \pi J \sigma_z^1 \sigma_z^2),
$
where $\omega_{1,2}/2\pi$ are the resonance frequencies of spins 1 and 2, 
$(\omega_{2} - \omega_{1})/2\pi =$ 226 Hz, $\omega_0 /2\pi$ is the 
spectrometer frequency ($\sim 400$ MHz), and $J$ is a scalar coupling 
strength (5.7 Hz).
The following on resonance ($\omega_0=\omega_1$) pulse sequence implements 
$V_T$ for the simulated period $\Omega T$:
\beqn
V_{T} = \left [\pi\right]_y^{1+2} - [\tau_1/2] -  \left [\pi\right]_y^{1+2} - 
[\tau_1/2+\tau_2].
\label{vtseq}
\eeqn
The symbol $[\pi]_y^{1+2}$ represents a $\pi$ angle radiofrequency pulse, 
oriented along the $y$ direction, on spins 1 and 2 (corresponding to the $V_i$) and $[\tau]$ represents a 
delay during which the 2-spin propagator $e^{-i{\cal H}_p^{0}\tau/\hbar}$ acts.
The time intervals are given by  $\tau_1 = \Omega T[1/(\pi J) - 2/(\omega_2-
\omega_1)]$ and $\tau_2= 2\Omega T/(\omega_2-\omega_1)$.

The experimental procedure is illustrated using $|s\rangle = |0\rangle + i|2
\rangle$ as follows:
\[
\ba{|c|}
\hline
\\
|s\rangle = |0\rangle \ +i|2\rangle \Leftrightarrow\; 
{ \tiny\acolspace{0.4mm}
|p\rangle\langle p| = \left(\ba{ccc} 
1 & \framebox[6mm]{ $ \ba{cc} 0 & 0 \ea $ } & -i\\
\framebox[5mm]{ $ \ba{c} 0 \\ 0 \ea $ } & 
\makebox[5mm]{ $ \ba{cc} 0 & 0 \\ 0 & 0 \ea $ }  & 
\framebox[5mm]{ $ \ba{c} 0 \\ 0 \ea $ }\\
i & \framebox[6mm]{ $ \ba{cc} 0 & 0 \ea $ } & 1\\
\ea\right)}
\\
\\
\;\stackrel{V_T}{\Longrightarrow}\; 
| p_T\rangle \langle p_T | = 
{ \tiny
\acolspace{1mm}
\left(\ba{ccc} 
1 & \framebox[6mm]{ $ \ba{cc} 0 & 0 \ea $ } & -i e^{i2\Omega T}\\
\framebox[5mm]{ $ \ba{c} 0 \\ 0 \ea $ } & 
\makebox[5mm]{ $ \ba{cc} 0 & 0 \\ 0 & 0 \ea $ }  & 
\framebox[5mm]{ $ \ba{c} 0 \\ 0 \ea $ }\\
i e^{-i2\Omega T} & \framebox[6mm]{ $ \ba{cc} 0 & 0 \ea $ } & 1\\
\ea\right)}
\\
\\
\ba{l}
\;\stackrel{{\rm Read \ } [\frac{\pi}{2}]_y^1}{\Longrightarrow}
\ea
{ \tiny
\acolspace{1mm}
\left(\ba{ccc} 
1 & \framebox[16mm]{ $  \ba{cc} 1 & i e^{i2\Omega T} \ea $ } & -i e^{i2\Omega T}
\\
\framebox[13mm]{ $ \ba{c} 1 \\ -i e^{-i2\Omega T} \ea $ } & 
\makebox[18mm]{ $ \ba{cc} 1 & i e^{i2\Omega T}\;\; \\ \;\;-i e^{-i2\Omega T} & 1 \ea $
 }  & 
\framebox[13mm]{ $ \ba{c} -i e^{i2\Omega T} \\ -1 \ea $ }\\
i e^{-i2\Omega T} & \framebox[16mm]{ $ \ba{cc} \;\;i e^{-i2\Omega T} & -1\;\; \ea $ } 
& 1\\
\ea\right)}\\
\\
\hline 
\ea
\]
The initial state $|p\rangle = \upup + i\dndn \leftrightarrow |s\rangle$, is 
easily prepared from the (pseudopure \cite{Cory}) state $\upup $. This in turn
 is produced from the thermal equilibrium state of 2,3-dibromothiophene by the
 sequence
$
\left [\pi/4\right]_x^{1+2} - \left [1/4J \right] - \left [\pi\right]_y^{1+2} -
 \left[1/4J \right]-  \left [-5\pi /6 \right]_y^{1+2} - \left[ G \right],
$
where the magnetic field gradient [$G$] destroys off-diagonal terms in the 
density matrix. The sequence \eqn{vtseq} for $V_T$ then leads to 
$|p_T\rangle\langle p_T|$. Since the simulated system should evolve coherently
 according to the difference in energy levels of the various superpositions, 
$|p_T\rangle\langle p_T|$ above expresses a $2\Omega T$ dependence.
NMR experiments are sensitive only to transverse dipolar magnetization, 
corresponding to the boxed components in the density matrices above. Thus a 
final read pulse is needed to rotate the $e^{\pm i2\Omega T}$ elements into 
view. The result manifests itself as a $2\Omega T$ oscillation of the spectral
peak heights as a function of the indirect dimension $T$.

The dynamics of the truncated QHO states $|0\rangle $, 
$(|0\rangle + i|2\rangle) $, and $(|0\rangle + |1\rangle + |2\rangle + 
|3\rangle) $ were simulated. Eigenstates like $|0\rangle$ do not evolve, as 
the simulation in Figure 1(a) shows.  Fig 1(b) shows the $2\Omega T$ 
oscillations discussed above for $|0\rangle + i|2\rangle $. In both Figs 1(a) 
and 1(b), $\left [\pi/2\right]_y$ read pulses were used. For 
$|s\rangle = |0\rangle + |1\rangle + |2\rangle + |3\rangle $, mixtures of  
$\Omega T$ and $3\Omega T$ oscillations can be observed in the spectra. For example, the operator $|0\rangle\langle 1 |$ corresponds to $\upup \updnd$ which is a transition of spin 2.  Thus the amplitude of the spin 2 peak will oscillate at $\Omega T$. In 
Fig 1(c) $\Omega T$ peak oscillations (on spin 2) are recorded while Fig 1(d) shows a superposition of $\Omega T$  and $3\Omega T$ (on spin 1). Since the two-spin system ($P$) 
has no natural triple quantum coherences, the latter coherence is entirely 
simulated. For Figs 1(c) and 1(d) read pulses were not required.

In general, scaling the above to include more levels will depend on the 
various couplings between the added spins. For larger spin systems certain 
couplings are small and therefore severely limit the time scale of the 
experiment. For the truncated QHO however, an effective hamiltonian that is free of all couplings results from mapping the energy eigenstate 
$|k\rangle$ to the spin eigenstate corresponding to the binary representation 
of $k$ in contrast to the Gray coding:
\[
\bar{\cal H}_p \;=\; \tfrac{1}{2}\hbar\Omega\left(2^n - \left[\sigma_z^1 + 
2\sigma_z^2 + 2^2\sigma_z^3 + \cdots+2^{n-1}\sigma_z^n\right]\right).
\]
This may be implemented by removing all scalar couplings and scaling all 
chemical shifts; for instance by methods analogous to ``chemical
shift concertina'' sequences introduced by Waugh  \cite{Waugh2}.

The Hamiltonian for an anharmonic oscillator,
$
{\cal H}_{\rm AHO} = 
\hbar\Omega[(\hat{N}+\tfrac{1}{2})+\mu(\hat{N}+\tfrac{1}{2})^2],
$
where $\mu$ is the anharmonicity parameter. 
The energy difference $\Delta E_m$ between the $m$th and $(m+1)$st energy 
level 
is 
$
\Delta E_m = \hbar\Omega [2\mu(m+1) +1 ].
$
Radiation at the frequency $\Delta E_m/h$ will drive a selective transition 
between these levels. The Hamiltonian for this selectively driven anharmonic 
oscillator is 
$
{\cal H}_{\rm AHO} +  \tfrac{1}{2}\hbar\Omega_R(|m\rangle\langle m+1| +  
|m+1\rangle\langle m|),
$ 
where $\Omega_R$ is the Rabi frequency.
Using the map \eqn{map1}, the $|0\rangle \leftrightarrow |1\rangle$ driven 
truncated Hamiltonian in particular maps to $\overline{\cal H}_p \equiv $
\[
\tfrac{1}{4}\hbar\Omega\left[\mu \sigma_z^1 - 4(4\mu+1)\sigma_z^2(1 + 
\tfrac{1}{2}\sigma_z^1) \right] + \tfrac{1}{4}\hbar\Omega_R\,\sigma_x^1(1 + 
\sigma_z^2).
\]
This is implemented on 2,3-dibromothiophene by the following pulse sequence
\beqn
V_{T} = [\tau_1/2] - \left [\pi\right]_y^{1} - [\tau_1/2] -  \left [3\pi /4 
\right]_y^{1} - [\tau_2] - \left [\pi /4 \right]_y^{1} .
\label{vtseq2}
\eeqn

For $\mu = -2/9$, and $\Omega_R = -2/9\Omega$, the time intervals are determined by $\Omega T/2\pi=9J\tau_2/(2\sqrt{2}) = (9/2)(m-\tau_1 \omega_2/2\pi) $, where {\it m} is an integer.  The receiver was set at $\omega_0/2\pi = \omega_1/2\pi - J/2$. Note that the map \eqn{map1} does not map the physical eigenstates to the 
simulated eigenstates (dressed states) of the driven oscillator emphasizing 
that knowledge of the eigenvalues/states of the simulated system is not 
assumed. Experimental results are shown in Figure 2.

In these studies, we have only considered unitary evolution and have explored 
the quantum dynamics for systems without dissipation. The decoherence 
\cite{Zurek} intrinsic in our physical system (characterized by the 
longitudinal and transverse magnetization relaxation times, $T_1$ and $T_2$) 
limits the time of the experiment. This then limits the number of periods that
can be simulated. Since the experimental ($t$) and simulated ($T$) time scales
need not be identified with each other, this can be interpreted as a 
restriction either on $\Omega$ or on $T$.  In Fig 2 the visible decay due to 
$T_2$ relaxation clearly shows this limitation. While decoherence can be 
controlled in principle by error correction \cite{shor1,steane,knill}, it 
would be difficult to utilize this in the weakly polarized physical system 
used here. Moreover, thermal equilibrium will not necessarily map to another 
configuration that is thermal. Decoherence, itself, may be simulated through 
specific non-unitary evolution; in NMR for example by magnetic field gradients
\cite{Cory2}. 

The aspects available in the simulation: controlled kinematics and dynamics, a
 driving field, and decoherence suggest a very general tool with which to study
other systems.

This work was supported in part by the U.S. Army Research office under
contract/grant number DAAG 55-97-1-0342 from the DARPA Ultrascale Computing
Program. R. L. thanks the National Security Agency for support.

\newpage

\begin{figure}
  \centering
  \epsfig{file=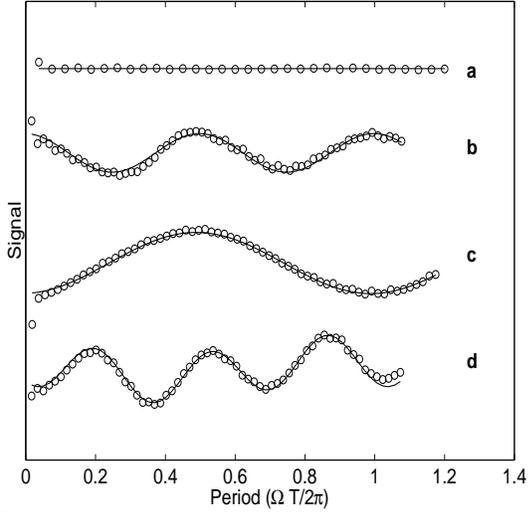,height=7cm,width=7cm}
  \caption{NMR peak signals from 2,3- dibromothiophene demonstrating a quantum 
simulation of a truncated harmonic oscillator as implemented by $V_{T}$ in 
\eqn{vtseq}. 
The various initial states express oscillations according to the 
energy differences between the eigenstates involved.  The solid lines are fits
 to theoretical expectations. {\bf a}, Evolution of the initial (pseudopure) 
state $|0\rangle$, showing no oscillation.  {\bf b}, Evolution of the initial 
state $|0\rangle + i|2\rangle$, showing $2\Omega $ oscillations.  {\bf c}, 
Evolution of the initial state $ |0\rangle + |1\rangle + |2\rangle + |3\rangle$
 showing the $\Omega $ and {\bf d} $3\Omega $ oscillations.  (Each trace in 
{\bf c} and {\bf d} is actually a combination of $\Omega $ and $3\Omega $ 
oscillations.) } 
  \label{hosc}
\end{figure}

\begin{figure}
  \centering
  \epsfig{file=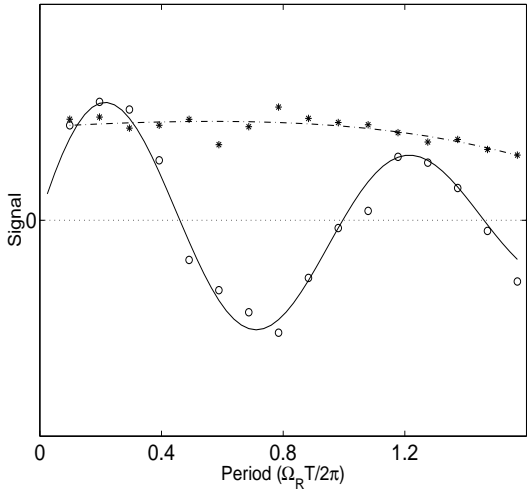,height=7cm,width=7cm}
  \caption{NMR peak signals from 2,3- dibromothiophene demonstrating a quantum 
simulation of a driven, truncated anharmonic oscillator as implemented by 
$V_{T}$ in \eqn{vtseq2}.  
When the (0,1) transition is selectively driven, the
 initial state $|0\rangle$  ($\circ$) undergoes Rabi ($\Omega_R$) oscillations
 to the $|1\rangle $ state, whereas the $|2\rangle $ state ($\ast$) does not 
evolve under the simulated Hamiltonian. The exponential decay due to natural 
decoherence in $P$ is clear.} 
  \label{ahosc}
\end{figure}

\end{document}